\documentclass[preprint,aip,jcp]{revtex4-1}
\usepackage{graphicx}
\usepackage{dcolumn}
\usepackage{bm}
\usepackage[version=3]{mhchem}

\begin{document}
\title{Gaussian-Charge Polarizable Interaction Potential for Carbon Dioxide}
\author{Rasmus A. X. Persson}
\email{rasmus.persson@chem.gu.se}
\affiliation{Department of Chemistry, University of Gothenburg, Sweden}

%\date{\today}

\begin{abstract}
A number of simple pair interaction potentials of the carbon dioxide molecule
are investigated and found to underestimate the magnitude of the second virial
coefficient in the temperature interval 220 K to 448 K by up to 20\%. Also the
third virial coefficient is underestimated by these models. A rigid,
polarizable, three-site interaction potential reproduces the experimental
second and third virial coefficients to within a few percent. It is based on
the modified Buckingham exp-6 potential, an anisotropic Axilrod-Teller
correction and Gaussian charge densities on the atomic sites with an inducible
dipole at the center of mass. The electric quadrupole moment, polarizability
and bond distances are set to equal experiment. Density of the fluid at 200 and
800 bars pressure is reproduced to within some percent of observation over the
temperature range 250 K to 310 K. The dimer structure is in passable agreement
with electronically resolved quantum-mechanical calculations in the literature,
as are those of the monohydrated monomer and dimer complexes using the
polarizable GCPM water potential. Qualitative agreement with experiment is also
obtained, when quantum corrections are included, for the relative stability of
the trimer conformations, which is not the case for the pair potentials.
\end{abstract} 


\maketitle

\section{Introduction}
For carbon dioxide (CO$_2$), being an important industrial chemical, numerous
interaction potentials (IPs) have been proposed, surpassed perhaps only by
water in the amount of computer attention it has attracted among the molecular
models. There are the parametric fits to the {\em ab initio} potential energy
surface of the dimer, such as the the IPs of Steinebrunner and
coworkers,\cite{steinebrunner98} of Bukowski and colleagues,\cite{bukowski99}
and of Bock {\em et al.} \cite{bock00} The most successful and widely used IPs,
however, have been fitted against bulk properties known from observation, such
as the vapor-liquid equilibrium \cite{potoff98,vrabec01,zhang05} (VLE) or the
crystal lattice parameters \cite{murthy81,*murthy83}, and still others against
experimental properties of the dilute gas, such as the second virial
coefficient. \cite{koide74,macrury76} One may ask why this is so, but the {\em
ab initio} IPs employ a great number of fitting parameters, not always of clear
physical origin and, even so, interfacing them with other IPs, for instance
when studying mixtures, is technically difficult because suitable combining
equations are not known for the many parameters. Moreover, this problem is not
unique to the {\em ab initio} IPs: also some empirical IPs use truncated series
expansions \cite{koide74, macrury76, vrabec01} for both angular and radial
functions in a way which makes it difficult to interface them with IPs of
radial site-site interactions.  Hence, such IPs can find little use outside
simulations of the neat liquid. On the other hand, great success has been
enjoyed by the simple site-site interaction formula of one Lennard-Jones
interaction center, and one atomic point charge, centered on every atomic
site\cite{murthy81, *murthy83, harris95, potoff98, zhang05}. Restricting
ourselves to the rigid models, these can all be regarded as descendants of the
original\cite{murthy81, *murthy83} IP due to Murthy, Singer and MacDonald
(MSM). These IPs have the very appealing property that they can be readily
interfaced (``mixed'') with existing force fields, many of which share the
exact same mathematical form.  

Nature is not kind enough, however, to allow such a simplified description of
the CO$_2$ molecule at no cost. Even though very good experimental agreement
for a wide variety of properties is obtained by the simple, rigid model with
three Lennard-Jones centers and one point quadrupole developed by Merker {\em
et al.},\cite{merker10} like the EPM-2 model,\cite{harris95} it suffers from
experimental disagreement in more ``basic'' properties such as the
carbon-oxygen bond distance. With respect to the VLE envelope, the most
successful such simple MSM-type model to date is the IP due to Zhang and Duan
\cite{zhang05} (ZD). Nevertheless, despite its high accuracy in this property,
I report in this Note that the microstructure of the dimer, and the temperature
dependence of the second, $B_2(T)$, and third, $B_3(T)$, virial coefficients
are of much poorer quality. It should be pointed out, also, that the results
presented in Ref.~\onlinecite{zhang05} have been called into
question.\cite{merker08}

One striking omission from the published CO$_2$ IPs is that of many-body
effects.  That the CO$_2$ molecule lacks an electric dipole moment may have
dissuaded investigators from looking in this direction, but in the work leading
up to this Research Note, extensive trials indicated that it is not possible
with a non-polarizable IP of MSM-type to simultaneously fit $B_2(T)$ while
keeping the experimental agreement of the VLE envelope, at least not if the
experimental bond distance and electric quadrupole moment are to be kept
intact. The decision was then made in favor of a polarizable IP, to be
described in this Note, but the extra cost that the high-resolution solution of
the electric field equations incurs, even for a single polarization site, make
simulations of the vapor-liquid coexistence prohibitively expensive for
parametrization purposes. Instead, the temperature-dependence of the fluid
density at 200 bar was used as a test for the many-body (concentrated phase)
and $B_2(T)$ for the two-body (dilute phase) properties. Further developments
then introduced a three-body dispersion interaction of Axilrod-Teller
type,\cite{axilrod43} and $B_3(T)$ as a parametrization target. As a test of
the validity of the IP, numerous other properties are calculated without input
into the parametrization procedure. The model introduced in this work goes by
the moniker of Gaussian Charge Polarizable Carbon Dioxide (GCPCDO). This is
because of its great similarity to the highly successful GCPM water
\cite{paricaud05} from which it borrows most of the essential equations.

This Note is organized as follows. First, in Section \ref{sec1} a description
of the mathematical form of the IP is given, and also the details of the
calculations and the targeted properties of the parametrization.  Then in
Section \ref{sec2}, the results are presented and discussed. Finally, a brief
recapitulation of the main points is given in Section \ref{sec3}.

\section{Model and computational details}
\label{sec1}
\subsection{Electrostatic interaction}
We compute the interaction between partial charge $q_\alpha$ and partial charge
$q_\beta$ at a separation of $r_{\alpha \beta}$, through the formula
\begin{eqnarray}
u_\mathrm{q}(r_{\alpha \beta}) = \frac {q_\alpha q_\beta} {r_{\alpha
\beta}} \eta(r_{\alpha \beta}, \tau_\alpha, \tau_\beta)
\label{eq:chgenergy}
\end{eqnarray}
Here $\eta(r_{\alpha \beta}, \tau_{\alpha}, \tau_{\beta})$ is a function that
assures that the electrostatic interactions remain finite at all separations;
for large $r_{\alpha \beta}$ it approaches unity. Physically, this corresponds
to charges distributed over a finite volume in space and like this we avoid the
singularity of the potential at zero charge separation. For point charges,
$\eta$ is identical to unity and Eq.~(\ref{eq:chgenergy}) reduces to the
classical Coulomb law. We choose the following form for the $\eta$-function,
\begin{equation}
\eta(r_{\alpha \beta}, \tau_\alpha, \tau_\beta) = \mathrm{erf} \left
(\frac {r_{\alpha \beta}} {\sqrt{2(\tau_\alpha^2 + \tau_\beta^2)}} \right )
\end{equation}
which corresponds the physical case of two Gaussian charge distributions of
standard deviations $\tau_\alpha$ and $\tau_\beta$ interacting with each
other.\cite{chialvo98} 
%The electric potential around the Gaussian charge
%$\alpha$ is then given by (letting $q_\beta \to 1$ and $\tau_\beta \to 0$),
%\begin{equation}
%\phi_\alpha(r) = \frac {q_\alpha} r \mathrm{erf} \left (\frac {r} {\sqrt{2}
%\tau_\alpha } \right )
%\end{equation}

For now, we shall not concern ourselves with the general case of many-body
interaction, as given by both ``static'' and ``dynamic'' electron correlation,
but exclusively take care of that ``static'' correlation from the average
electric field around each molecule, {\em i. e.} electronic induction effects.
Furthermore, we do not carry this analysis beyond the dipole induction, {\em i.
e.} we consider only the gradient of the electric potential. An induced dipole
is hence positioned at the center of mass of molecule $l$ and is given by
\begin{equation}
\vec \mu_l = \vec E_l \left ( \alpha_{\bot} + |\cos(\theta)| (\alpha_\parallel
- \alpha_{\bot}) \right ) 
\end{equation}
where $\vec E_l$ is the electrostatic field at that point, $\alpha_\bot$ the
polarizability perpendicular to the molecular axis and $\alpha_\parallel$ that
parallel to the same. Both $\alpha_\bot$ and $\alpha_\parallel$ are known from
theory \cite{maroulis03} and their average agree in magnitude with that
ascertained in experiment, \cite{bridge66, chrissanthopoulos00} but the
interpolation between them has been chosen merely for convenience. $\theta$ is
the angle between the molecular axis and the direction of $\vec E_l$, the
electric field at site $l$. Because of the uncertainty in the precise form of
the polarization matrix, and the approximations involved with the rigid rotor,
the polarizabilities have been rounded to only two significant digits. In its
turn, the electric field is computed as the sum of the contributions due to the
permanent charges and that due to the other induced dipoles,
\begin{equation}
\vec E_l = \vec E_l^\mathrm q + \vec E_l^\mathrm \mu
\end{equation}
where $\vec E_l^\mathrm q$ is the electric field at the center of molecule $i$
due to the all the charges on the other molecules, and $\vec E_l^\mathrm \mu$
is the electric field at the same point, but due to all the other induced
dipoles. These are given by
\begin{equation}
\vec E_l^\mathrm q = \sum_{m \neq l} \sum_{\beta \in m} \frac {q_{\beta} (\vec
r_l - \vec r_\beta)} {r_{l\beta}^3} \left (\eta(r_{lm}, \tau_l, \tau_\beta) -
\frac {\partial} {\partial r_{lm}} \eta(r_{lm}, \tau_l, \tau_\beta) \right )
\end{equation}
and
\begin{equation}
\vec E_l^\mathrm \mu = \sum_{m \neq l} \mathbf T_{lm} \vec \mu_m
\end{equation}
where the tensor $\mathbf T_{lm}$ is obtained from the Hessian of equation
(\ref{eq:chgenergy}), with -- following Paricaud {\em et al.} \cite{paricaud05}
-- the charge width parameter of the molecular center equal to that of the
center atom.  If the dipoles are converged, the extra energy of interaction due
to the polarization is given by, \cite{ahlstrom89}
\begin{equation}
U_\mathrm{pol} = -\frac 1 2 \sum_{l=1}^N \vec \mu_l \cdot \vec E_l^\mathrm{q}
\end{equation}

\subsection{Dispersion interaction}
First of all, the modified Buckingham exp-6 potential is adopted between atoms
$\alpha$ and $\beta$, 
\begin{equation}
u_{\mathrm{disp},2}(r_{\alpha \beta}) = \frac {\epsilon_{\alpha \beta}} {1 - 6
/ \gamma_{\alpha \beta}} \left [\frac {6} {\gamma_{\alpha \beta}} \exp \left \{
\gamma_{\alpha \beta} \left (1 - \frac r \sigma_{\alpha \beta} \right ) \right
\} - \left ( \frac \sigma r \right )^6 \right ]
\end{equation}
This represents the pairwise additive part of the dispersion interaction as
well as the steric repulsion between atoms. Also for this interaction, a
hard-core is introduced at $r_{\alpha \beta} = 0.57 \sigma_{\alpha \beta}$ to
avoid the spurious behavior of this potential at short range. This is the same
hard-core cutoff used by Paricaud and coworkers \cite{paricaud05} in the GCPM
water model. Investigations indicated that the results are not very sensitive
to the shortening of this hard-core radius to $0.35 \sigma_{\alpha \beta}$, but
the speed of simulation is. Here $\epsilon_{\alpha \beta}$, $\gamma_{\alpha
\beta}$ and $\sigma_{\alpha \beta}$ are atomic interaction parameters, related
to the well-depth, steepness and position, respectively, of the dispersion
interaction. Second, a modified Axilrod-Teller term is added for all molecular
triples. This is the triple-dipole dispersion correction to the van der Waals
interaction which was first published by Axilrod and Teller. \cite{axilrod43}
In the derivation by Axilrod,\cite{axilrod51} spherical polarizabilities are
assumed, but for anisotropically polarizable molecules, such as CO$_2$, his
result does not hold. To investigate this case, we quickly recapitulate
Axilrod's derivation\cite{axilrod51} where we dispose of the assumption of
spherical polarizabilities. 

For each of the three molecules $l$, $m$ and $n$, we define a local coordinate
system where the $z$-axis is normal to the plane of the molecular centers, the
$x$-axis parallel to the bisector of the angle spanned by the two other
molecules and the $y$-axis mutually orthogonal to the $x$- and $z$-axes.
Hence, the $z$-axes all coincide between the three local coordinate systems,
but the $x$- and $y$-axes need not.  Assume now that the electronic structure
of each molecule is independent and described by a wavefunction that factorizes
into separable $x$-, $y$- and $z$-contributions in its local coordinate system,
{\em i. e.} for molecule $l$,
\begin{equation}
\psi_l(x_l, y_l, z_l) = X(x_l) Y(y_l) Z(z_l)
\end{equation}
We write the third-order perturbation correction to the groundstate energy,
$W_0'''$, which in Axilrod's notation is (Eq.~[5a] in
Ref.~\onlinecite{axilrod51}),
\begin{equation}
W_0''' = \mathop{\sum_{j}}_{j \neq k, 0} \mathop{\sum_{k}}_{k \neq 0}
\frac{H_{0j}' H_{jk}' H_{k0}'} {(W_j^0 - W_0^0)(W_k^0 - W_0^0)} 
\end{equation}
Here $H'_{jk}$ is the matrix element of the dipole perturbation for states $j$
and $k$, and $W_j^0$ is the energy of state $j$. Invoking the closure
approximation, \cite{atkins05} this equation may be approximately recast as
\begin{equation}
W_0''' \approx \frac 1 {\nu'} \mathop{\sum_{j}}_{j \neq k, 0}
\mathop{\sum_{k}}_{k \neq 0} {H_{0j}' H_{jk}' H_{k0}'} \equiv
u_{\mathrm{disp},3}(l, m, n) \label{pert}
\end{equation}
where
\begin{equation}
\nu' = \langle (W_j^0 - W_0^0)(W_k^0 - W_0^0) \rangle_{jk}
\end{equation}
and $\langle \ldots \rangle_{jk}$ signifies the arithmetic average over $j$ and
$k$. Eq.~(\ref{pert}) serves to define the three-body correction to the
dispersion energy that we will use. 

Formally, the set of matrix elements $\{H_{jk}'\}$ covers all possible excited
states but we shall assume contributions to the sum only from the first excited
orbital of each symmetry for each molecule. That is, the three lowest excited
states of the arbitrary molecule $l$ are assumed to be $X^*(x_l) Y(y_l)
Z(z_l)$, $X(x_l) Y^*(y_l)
Z(z_l)$ and $X(x_l) Y(y_l)
Z^*(z_l)$ where the asterisk denotes the next higher-energy orbital. Because
they share a common orthogonal $z$-axis, only mixed excited states for $x$- and
$y$- components between the molecules contribute to the sum over states. Hence,
the matrix elements for the sequences of possible excitations are exhaustively
given in Table I of Ref.~\onlinecite{axilrod51} and the explicit form of these
matrix elements is provided in Eq.~(29) of the same reference, except that the
common factor  $M^2$ is no longer applicable because the transition diople
moment is no longer the same for the different components. Instead, given the
two molecules $l$ and $m$, excited in their $x$ and $y$ orbitals,
respectively, we write the corresponding matrix element ({\em cf.} Eq.~[29] in
Ref.~\onlinecite{axilrod51})
\begin{equation}
(x_m, y_l) =  (M_{x_m} M_{y_l} / (4 \pi \varepsilon R_{ml}^3))(2 \cos \frac 1
2 \gamma_m \sin \frac 1 2 \gamma_l - \sin \frac 1 2 \gamma_m \cos \frac 1 2
\gamma_l) \label{eq:matel}
\end{equation}
where $M_{x_m}$ denotes the expectation value of transition dipole moment along
the $x$-axis of molecule $m$ and $\gamma_m$ the angle defined by the molecules
$l,m$ and $n$ with its the apex in molecule $m$. All the other matrix elements
follow by analogy with Ref.~\onlinecite{axilrod51}. We have written
Eq.~(\ref{eq:matel}) in  a general form with $\varepsilon$ being the
permittivity of the medium in which the molecules are dispersed. It is most
reasonable to take this as the permittivity of free space. Since because of the
very short-range nature of the interaction (it tapers off as the inverse ninth
power of distance) it is unreasonable to assume that a homogeneous medium of
CO$_2$ molecules can be accommodated between the interacting molecules. The
final approximation is to replace the $M$-factors  by the square-root of the
corresponding polarizabilities. Collecting the constants of proportionality in
the common prefactor of Eq.~(\ref{pert}), which is then seen to have dimensions
of reciprocal energy, we treat it as a fitting parameter and denote it by $1 /
\nu$ proper.  

With anisotropic polarizabilities, the sum over states in Eq.~(\ref{pert}) does
not simplify to the simple form given in the original references
\cite{axilrod43, axilrod51} and the complicated closed-form expression will not
be reproduced here. In any case, since it involves  sines and cosines it is not
optimal from a computational point of view; it is much more efficient in terms
of the total number of floating-point operations to calculate the truncated sum
over states directly. To this end, the half-angle formulas are used to
rewrite the matrix elements, such as the one in Eq.~(\ref{eq:matel}), in terms
of dot products and square-roots, which are much more efficient in terms of CPU
cycles than trigonometric functions. The polarizabilities are calculated, like
before, as the interpolation
\begin{equation}
\alpha(\theta) = \alpha_\bot + |\cos(\theta)|(\alpha_\parallel - \alpha_\bot)
\end{equation}
where $\theta$ is, once again, the angle to the molecular axis. The limiting
polarizabilities are taken to be the same as the static ones.

In total, after self-consistent solution of the induced dipoles, the energy of
interaction among $N$ molecules is given by
\begin{equation}
U = \sum_{l=1}^N\sum_{m>l}^N \sum_{\alpha \in l} \sum_{\beta \in m}
\left (u_\mathrm{q} (r_{\alpha \beta}) + u_{\mathrm{disp},2}(r_{\alpha \beta})
\right ) - \frac 1 2 \sum_{l=1}^N \vec \mu_l \cdot \vec E_l^\mathrm q  +
\sum_{l=1}^N \sum_{m>l}^N \sum_{n>m}^N u_{\mathrm{disp},3}(l,m,n) 
\end{equation}
The analytical gradient of this expression is very involved, with the
chain-rule giving factors proportional to the gradient of the polarizability.
Consequently, when the gradient has been needed, for instance, in energy
minimization, it has been calculated numerically using the finite-difference
approximation.

\subsection{Parametrization strategy}
\subsubsection{Gas-phase properties}
A number of parameters have not been optimized, but simply assigned from
plausible experimental values in the literature. Thus, the bond length is fixed
at 1.161 \AA, midway between published values of 1.160 \AA\ and 1.162 \AA\ by
experimental groups,\cite{harmony79, granar86} and the partial charges on the
atoms are chosen to reproduce the experimental quadrupole
moment.\cite{buckingham63,harries70} To further reduce the number of free
parameters, the charge width of the oxygen atom, $\tau_\mathrm O$, was set
equal to 0.610 \AA, the value of the oxygen atom in GCPM water
,\cite{paricaud05} and, the corresponding quantity for carbon was scaled
according to the ratio of the $\sigma$ parameters ({\em vide infra}). These
parameters were not optimized.  Moreover, I introduced the additional
constraint of $\gamma_{\alpha \alpha} = \gamma_{\beta \beta} = \gamma_{\alpha
\beta}$ and for the remaining parameters, the following ``mixing rules'' were
adopted for unlike interactions of the modified Buckingham exp-6 potential,
\begin{eqnarray}
\epsilon_{\alpha \beta} & = & \frac {2 \epsilon_{\alpha \alpha} \epsilon_{\beta
\beta}} {\epsilon_{\alpha \alpha} + \epsilon_{\beta \beta}} \label{eqn:mix1} \\
\sigma_{\alpha \beta} & = & \frac 1 2 \left ( \sigma_{\alpha \alpha} +
\sigma_{\beta \beta} \right )
\label{eqn:mix2}
\end{eqnarray}
Eq.~(\ref{eqn:mix1}) can be justified with appeal to the London
formula,\cite{eisenschitz30} in which the harmonic average of the ionization
energy is taken. Normally, it is the geometric average of the polarizabilities
in said equation that lends its mathematical form to the combining rule for the
$\epsilon$-parameters. However, the precise form of the mixing rules are of a
secondary concern, as they serve mainly to reduce the parameter space that has
to be fit, and provided the atomic interaction parameters do not turn out to be
very different from each other, the result will be insensitive to reasonable
choices of mixing rules.

A test set of potentials with predefined $\sigma$ values, covering a broad
range, but with the ratio, $\sigma_\mathrm C / \sigma_\mathrm O$ of the carbon
sigma value, $\sigma_\mathrm C$, and the oxygen sigma value, $\sigma_\mathrm
O$, conserved at $1.0483$ were investigated. This value was arbitrarily chosen
early in the development of the model and never subjected to revision. For
each such class of IPs, manual tuning of the $\epsilon$-parameters in
trial-and-error fashion was made to obtain a good fit for $B_2(T)$ against
experimental data and a reasonable binding energy and geometry of the dimer
structure. $B_2(T)$ was calculated from the Mayer sampling method
\cite{singh04} over $5 \times 10^8$ Monte Carlo steps. Typically, with this
number of Monte Carlo steps, the resulting uncertainty, calculated from the
block average method \cite{flegal07} with blocks of $10^5$ cycles, is of the
order of one percent.  The induced dipoles were iteratively solved to within a
tolerance of $3 \times 10^{-10}$ D at all times. In the course of this
parametrization, it was found that the $\gamma$-parameter plays a most crucial
role. This parameter controls the steepness of the modified Buckingham
potential, without affecting well depth or location. Small values lead to a
flat minimum and $B_2(T)$ turned out to be very sensitive to this parameter.
Thus, it was found that, to fit $B_2(T)$, this parameter had to be increased
from its initial estimate of 12.75 (taken from GCPM water \cite{paricaud05}) to
15.50.    

Later in the development of the model, it was found necessary, with respect to
the liquid densities, to include also the three-body dispersion in the energy
expression. Because the $\nu$-parameter is completely independent of all dimer
properties, including $B_2(T)$, it was fitted independently so that $B_3(T)$
coincided with the data of Dushek {\em et al.}, \cite{dushek90} It was not
possible to reproduce, with this single parameter, also the data of Holste {\em
et al.},\cite{holste87} something which lends credence to the former
measurements. The value obtained, $\nu = 2.82 \times 10^4$ K, is reasonable in
that it is of the same order of magnitude as the Axilrod-Teller coefficient
obtained for argon, for which this coefficient is $2.1 \times 10^4$ K in the
appropriate units.\cite{leonard75}

\subsubsection{Bulk simulations}
The fluid densities were extracted from isothermal-isobaric Metropolis Monte
Carlo simulations \cite{metropolis53} for an ensemble of 200 molecules in
periodic boundary conditions of cubic symmetry over $M = 2.5 \times 10^7$
steps, run in parallel over five independent Markov chains. Standard errors of
the mean were estimated from the block average method \cite{flegal07} with
$\sqrt{M}$ blocks.

For the modified Buckingham potential, with neglect of interactions beyond the
cutoff the energy was corrected by 
\begin{equation}
U_\mathrm{LRC} = u_\mathrm{LRC}^\mathrm{CC} + 4 u_\mathrm{LRC}^\mathrm{CO} + 4
u_\mathrm{LRC}^\mathrm{OO}
\end{equation}
where $u_\mathrm{LRC} = \frac \rho 2 \int_{r_\mathrm{c}} ^{\infty} 4 \pi r^2
U_\mathrm{exp-6}(r) \mathrm d r$ and superscripts indicate between which two
atom types the interaction is computed. Here $\rho$ is the number density of
molecules. The integral in question can be analytically computed, which yields
\begin{equation}
u_\mathrm{LRC}  =  \frac {2 \pi \epsilon \rho} {1 - 6/\gamma} \left [ \frac {6
\sigma} {\gamma^4} \left (r_c^2 \gamma^2 + 2 r_c^2 \gamma \sigma + 2 \sigma^2
\right ) \exp \left (\gamma - \frac {r_c \gamma} {\sigma} \right ) - \frac
{\sigma^6} {3 r_c^3} \right ]
\end{equation}
For the Axilrod-Teller terms, in light of their very strong distance
dependence, no long-range correction was deemed necessary.

For the electrostatic interaction, a long-range correction was introduced for
the induced dipole using the Onsager expression for the dipole reaction
field\cite{onsager36} with the dielectric constant taken from the
Clausius-Mosotti equation fitted against experimental measurements of the
dielectric constant.\cite{keyes30} This may seem to be very approximate but was
done for two principal reasons. First, lattice summation techniques such as the
Ewald sum \cite{frenkel01} introduce assumptions of periodicity which are
unsuitable for the uncorrelated nature of the molecules in a disordered phase.
Second, the quadrupolar interaction terms decay sufficiently fast with distance
so as to make their sum absolutely convergent. On the assumption of
uncorrelated molecules beyond the cutoff radius, the long-range correction
vanishes. Last but not least, the exact same procedure was used by Paricaud
{\em et al.} in their simulations of GCPM water.\cite{paricaud05}  As a
technical side-note, it should be pointed out that Ewald summation for Gaussian
charges is complicated by the fact that the required Fourier transform for the
reciprocal space sum is not analytically tractable.  Using the reciprocal space
sum for point charges, with due corrections in direct space and a long cutoff
radius, is an alternative but inefficient. 

Each trial move consisted of randomly displacing and rotating from one up to
four molecules. Interaction cutoffs were introduced at half the box length. For
the three-body dispersion, this was interpreted to mean that all interacting
molecules had to be within cutoff of each other. Simple mixing \cite{eyert96}
with a mixing factor of 10\% was used to enforce and speed convergence. Because
of increased computational load, the induced dipoles were solved to within a
tolerance of $3.4 \times 10^{-5}$ D per molecule. This is still less tolerant
than many other reported simulations of polarizable IPs: Ren and Ponder
\cite{ren03} report simulations on liquid water with their AMOEBA force field
where the dipoles are solved to within $10^{-2}$ D precision; Paricaud {\em et
al.}.\cite{paricaud05} solve the induced dipoles to within $5 \times 10^{-5}$ D
in their simulations of GCPM water. However, further relaxation of the
tolerance is unnecessary as the greater part of the simulation time is not
spent on the self-consistent solution of the elctrostatic field equations, but
(about 70\%) on the evaluation of $\sum_{lmn} u_{\mathrm{disp},3}(l,m,n)$. This
computation can be significantly sped up with shorter interaction cutoffs.

Two types of bulk simulations were performed. A series of $NpT$-simulations to
determine the density and constant-pressure heat capacity of the model fluid
and $NVT$-simulations to determine the radial distribution functions of the
model fluid. Together with the calculation of the virial coefficients, these
served to parametrize the model. All parameters except for $\nu$ were decided
for the potential model lacking the $u_3$ term, which has no effect on either
the dimer binding energy, its geometry or the second virial coefficient.
Selection among candidates of this pairwise dispersion model was effected by
comparing densities from the $NpT$-simulation with experiment, although the
long-range correction was not fully developed at this time. Subsequent
introduction of the long-range correction lead to an increase of the computed
densities, later corrected by the introduction of the three-body dispersion
interaction.  The final parameters, extracted from these tests, are listed in
Table \ref{tab:param}.
\begin{table}
\caption{Interaction parameters and experimental observables for the GCPCDO
model and, where appropriate, experimental reference values.}
\footnotetext[1]{Ref. \onlinecite{bridge66}} 
\footnotetext[2]{Ref. \onlinecite{harmony79}} 
\footnotetext[3]{Ref. \onlinecite{granar86}}
\footnotetext[4]{Ref. \onlinecite{buckingham63}}
\footnotetext[5]{Ref. \onlinecite{harries70}}
\label{tab:param}
\begin{ruledtabular}
\begin{tabular}{c r r} 
Parameter  & GCPCDO & Exp. \\ \hline
$\sigma_\mathrm O$ / \AA & $3.347$ &       \\ 
$\sigma_\mathrm C$ / \AA & $3.193$ &       \\ 
$\epsilon_\mathrm O$ / K & $67.72$ &       \\ 
$\epsilon_\mathrm C$ / K & $71.34$ &      \\ 
$\nu$ / K                & $2.82 \times 10^4$ \\
$\gamma_\mathrm O$ / \AA & $15.50$ &       \\ 
$\gamma_\mathrm C$ / \AA & $15.50$ &      \\ 
$\tau_\mathrm O$ / \AA & $0.6100$ & \\
$\tau_\mathrm C$ / \AA & $0.5819$ & \\
$q_\mathrm O$ / $e$ & $-0.3321$ &       \\ 
$q_\mathrm C$ / $e$ & $0.6642$ &       \\ 
$\alpha_\bot$ / \AA $^3$ & $1.95$ & $1.929$\footnotemark[1]
\\ 
$\alpha_\parallel$ / \AA $^3$ & $4.05$ &
$4.038$\footnotemark[1] \\ 
Bond length / \AA & $1.161$ & $1.160$\footnotemark[2] \\
                  &       & $1.162$\footnotemark[3] \\ 
Quadrupole / (e \AA$^2$) & $-0.90$ & $-0.85$\footnotemark[4] \\ 
                                      &     & $-0.90$\footnotemark[5] \\
				      &     & $-0.96$\footnotemark[5] \\
\end{tabular}
\end{ruledtabular}
\end{table}

\section{Results and discussion}
\label{sec2}
\subsection{Virial coefficients}
Not only have calculations for the parametrization of the GCPCDO IP been
carried out, but in the investigations leading up to its formation, I also
investigated some other CO$_2$ IPs. The VLE envelopes of these models have all
been thoroughly investigated in Ref.~\onlinecite{zhang05}, in which it was
established that the ZD potential exhibits near-perfect experimental agreement
in this property. While this excellent agreement might not be reproducible in
every aspect,\cite{merker08} the critical temperature and density cannot lie
very far from their experimental counterparts nonetheless, because the relative
deviations between the coexistence densities reported in Ref.
\onlinecite{zhang05} and Ref. \onlinecite{merker08} decrease and approach the
experimental values when nearing the experimental critical temperature.
Moreover, the variation in $B_2(T)$ is not very great between the models;
except possibly for the TraPPE IP \cite{potoff01} which---it is interesting to
point out---is that of the non-polarizable IPs of MSM-type, which exhibits the
best experimental agreement for both $B_2(T)$ and $B_3(T)$. For the GCPCDO IP,
having had this as one of its goals in the parametrization, the fit is very
good with the relative error never exceeding 5\%. The results are reported for
a select number of temperatures in Table
\ref{tab:b2}. 
\begin{table}
\caption{Second virial coefficients for the ZD, MSM, EPM2, TraPPE and GCPCDO
IPs, as well as experimental results, at a select number of
temperatures. For the computed results, bracketed numbers indicate the
estimated standard error of the mean in the last digit from the Mayer
sampling\cite{singh04} Monte Carlo integration.}
\label{tab:b2}
\begin{ruledtabular}
\begin{tabular}{l r r r r r r} 
& \multicolumn{3}{l}{$B_2$ / (cm$^3$ mol$^{-1}$)}  & & & \\
$T$ / K & ZD           & MSM         & EPM2       & TraPPE     & GCPCDO & Exp. \\\hline
$220.00$ & $-200.6(6)$ & $-204.7(6)$ & $-206.6(7)$ & $-222.0(7)$ & $-247.2(9)$
								  &
								  $-247.50$\footnotemark[1] \\ &          &         &          &           &              &
       $-247.52$\footnotemark[2]\\
$240.00$ & $-168.1(5)$ & $-171.5(6)$ & $-170.9(6)$ & $-183.5(6)$  & $-202.8(9)$
& $-202.83$\footnotemark[1] \\
       &         &         &          &           &           &
       $-202.13$\footnotemark[2] \\ 
$260.00$ & $-141.7(5)$ & $-145.0(5)$ & $-144.7(5)$ & $-153.7(6)$  & $-169.8(9)$
& $-168.92$\footnotemark[1] \\
       &         &         &          &           &           &
       $-168.27$\footnotemark[2] \\
$280.00$ & $-120.6(5)$ & $ -123.3(5)$ & $-122.1(5)$ & $-129.3(5)$ & $-143.7(7)$
& $-142.70$\footnotemark[1] \\
       &           &         &         &          &           &
       $-142.11$\footnotemark[2] \\
$300.00$ & $-104.4(4)$ & $-106.2(4)$ & $-105.0(4)$ & $-110.3(5)$ & $-123.5(6)$
& $-121.70$\footnotemark[1] \\
         &          &          &       &          &          & $-121.35$\footnotemark[2] \\
$340.00$ & $-78.0(4)$ & $ -79.5(4)$ & $-78.0(4)$ & $-81.9(4)$ & $-91.3(4)$
&   $-90.57$\footnotemark[2]  \\
$423.15$ & $-43.5(3)$ & $-44.8(3)$ & $-43.8(3)$ & $-45.6(4)$ & $-52.2(4)$ &
$-51.25$\footnotemark[1] \\
$448.15$ & $-36.6(3)$  & $-37.5(3)$ & $-36.5(3)$  & $-37.7(3)$ & $-43.9(3)$
& $-43.51$\footnotemark[1]  \\ 
\end{tabular}
\end{ruledtabular}
\footnotetext[1]{Ref. \onlinecite{holste87}}
\footnotetext[2]{Ref.  \onlinecite{dushek90}}
\end{table}

It does seem surprising that IPs that fare very well in reproducing VLE
properties should fail so remarkably at reproducing the much ``simpler''
property $B_2(T)$. If $B_2(T)$ is overestimated, then
a compensating underestimation of $B_3(T)$ seems very likely. Direct
calculation of $B_3(T)$ for the computer models seem to confirm this.  However,
contrary to the case of $B_2(T)$, good quality experimental data are very hard
to find for $B_3(T)$.  Different authors report widely different results, over
the same temperature range. In the narrow range around the critical
temperature, however, both Holste {\em et al.} \cite{holste87} and Dushek {\em
et al.} \cite{dushek90} report measurements which are in at least slight mutual
concordance. In Table \ref{tab:b3}, these measurements are reported and
compared with predictions from the IPs. As is evident, the pairwise additive IPs
underestimate $B_3(T)$ across the whole temperature range. 
\begin{table}
\caption{Third virial coefficients for the ZD, MSM, EPM2, TraPPE and GCPCDO
IPs, as well as experimental results, at a select number of
temperatures. For the computed results, bracketed numbers indicate the
estimated standard error of the mean from the Mayer sampling \cite{singh04}
Monte Carlo integration.} \label{tab:b3}
\begin{ruledtabular}
\begin{tabular}{l r r r r r r} 
& \multicolumn{3}{l}{$B_3$ / (cm$^6$ mol$^{-2}$)}  & & & \\
$T$ / K & ZD      & MSM       & EPM2 & TraPPE   & GCPCDO & Exp. \\ \hline
$280.00$ & $2920(30)$ & $2940(30)$   & $2910(30)$ & $3080(40)$ & $5140(60)$  &
$5636$\footnotemark[1] \\
       &          &           &        &         &           &
       $5165$\footnotemark[2] \\ 
$300.00$ & $2820(20)$ & $2860(20)$ & $2922(20)$ &  $3060(30)$ & $4790(60)$   &
$4927$\footnotemark[1] \\
       &           &         &       &          &            &
       $4753$\footnotemark[2] \\ 
$320.00$ & $2690(20)$ & $2760(20)$ & $2700(20)$ & $2870(20)$  & $4460(50)$   & 
$4423$\footnotemark[1] \\
       &           &         &       &          &            &
$4360$\footnotemark[2]\\
$340.00$ & $2530(20)$ & $2570(20)$ & $2560(20)$ & $2680(20)$  & $4046(50)$   &
$3996$\footnotemark[2] \\
\end{tabular}
\end{ruledtabular}
\footnotetext[1]{Ref. \onlinecite{holste87}} 
\footnotetext[2]{Ref. \onlinecite{dushek90}}
\end{table}

\subsection{Volumetric properties}
To investigate the properties of the many-body potential with more than just
three bodies, the density and the heat capacity at constant pressure, both
readily extracted from the $NpT$ simulations, serve as indicators. These results
are summarized in Table \ref{tab:rho} with experimental data. It is
surprising that the agreement with experiment is considerably better at the
higher densities. This increasing discrepancy between theory and experiment can
be tentatively attributed to the fourth virial coefficient, $B_4(T)$. It is
clear that in order to explain these results, $B_4(T)$ (or possibly higher
virial coefficients) must rise quicker with temperature for the GCPCDO model
than for its experimental counterpart. Unfortunately, quality experimental
values of $B_4(T)$ are not known but it is worth pointing out that if one
applies the virial equation of state truncated after $B_3(T)$, the computed
densities at 290 K and 310 K turn out to be $0.915 \pm 0.018$ g / cm$^3$ and
$0.791 \pm 0.015$ g / cm$^3$, respectively, at 200 bar pressure. These values
are indeed very close to the simulated values reported in Table \ref{tab:rho}
and indicate that $B_4(T)$ is overestimated and close to zero. However, because
at the higher pressure, the truncated virial equation of state predicts
densities {\em higher} than the simulated ones, it is clear that higher-order
coefficients must be compensating for errors in $B_4(T)$. This is not
surprising, at high densities, the steric effects become the dominant mode of
interaction. Sadly, the precise causes of these deviations in the virial
coefficients is unknown. Unlike the other models investigated in this work, we
have come some way in correcting $B_2(T)$ and $B_3(T)$ to their correct
experimental values. However, we may safely say that the interactions of the
CO$_2$ molecule are not as simple as one may at first suppose.

Also shown in Table \ref{tab:rho} is the constant-pressure heat capacity which
was calculated from the fluctuation formula, \cite{hill56}
\begin{equation}
C_\mathrm p = \frac 1 {N k T^2} \left (\langle H^2 \rangle - \langle H \rangle^2
\right ) 
\end{equation}
where $H = U + p V + 5 k T / 2$ is the enthalpy, the last term being the
classical kinetic contribution of a linear rigid body and $k$ the Boltzmann
constant, $p$ the pressure, $U$ the potential energy, $V$ the volume, $N$ the
number of molecules and $T$ the temperature. For a completely fair comparison with the experimental values, also the internal
vibrational degrees of freedom should be included. Assuming harmonic behavior,
for each normal mode of frequency $\nu$ this quantized harmonic contribution to
the heat capacity is then 
\begin{equation}
C_\mathrm{p,h} = k \left (\frac {h \nu} {k T} \right )^2 \left (\exp
\left ( \frac {h \nu} {k T} \right ) - 1 \right )^{-2} \exp \left (\frac {h
\nu} {k T} \right )  \label{eq:hcapcor}
\end{equation}
where $h$ is the Planck constant. Taking into account the experimental
frequencies \cite{martin32,ouazzany87} of the four harmonic normal modes of the
CO$_2$ molecule, this term is added to $C_\mathrm{p}$ and reported as the
corrected values in Table \ref{tab:rho}.  
% The normal modes of the carbon dioxide molecule have wavenumbers 667, 1285
% and 2349 cm^-1. The first one is doubly degenerate.
Not surprisingly, this expression compares favorably with the experimental
$C_\mathrm p$ extrapolated to vanishing density.\cite{webbookfluid} For
instance, the experimentally ascertained intramolecular contribution is 5.74 J
/ (K mol) at 250 K, whereas from Eq.~(\ref{eq:hcapcor}) one has 5.78 J / (K
mol). At 310 K, the experiments indicate 8.58 J / (K mol) and from
Eq.~(\ref{eq:hcapcor}) we have 8.68 J / (K mol). It is computationally too
demanding, at present, to include the quantized vibrations in the bulk
simulation of the GCPCDO IP and the approximation of separable internal and
external degrees of freedom is expected to be fair. 

The general overestimation of the heat capacity, even before the correction for
intramolecular degrees of freedom, is due, at least in part, to the assumption
of classical translational and rotational degrees of freedom. Especially at
high density, free rotation and translation is not possible and the rotational
and translational degrees of freedom are in effect partly quantized librational
modes. Unlike the intramolecular degrees of freedom, these are highly coupled
and there exists no viable computational approximation for their contribution
to the heat capacity. The very large overestimation of the heat capacity at 310
K and 200 bar cannot, however, be attributed to this effect alone. Moreover, at
this state point the discrepancy in density between real CO$_2$ and GCPCDO is
so large that a more fair comparison (as relates to $C_\mathrm p$) is with the
experimental $C_\mathrm p$ at $0.79$ g / cm$^3$ density, which is
\cite{webbookfluid} $119$ J / (K mol), not too far off from the computed value. 

\begin{table}
\caption{Calculated and experimental fluid properties at 200 and 800 bar
pressure and temperatures of 250 to 310 K. Heat capacities assume classical
contribution of $3k$ for the rotational and translational degrees of freedom.
The corrected heat capacities include the heat capacity of the quantized
internal degrees of freedom within the harmonic approximation. For the
calculated results, numbers in parentheses indicate the estimated standard
error of the mean in the last digit.}
\label{tab:rho}
\begin{ruledtabular}
\begin{tabular}{ll lr lcr} 
$p$ / bar & $T$ / K & \multicolumn{2}{l}{$\rho$ / (g / cm$^3$)} &
\multicolumn{3}{l}{$C_p$ / (J / [K mol])} \\
   &       & Calc. & Exp.\footnotemark[1] & Calc. & Corr. & Exp.\footnotemark[1]
   \\ \hline
200 & 250.0 & 1.100(4) & 1.105 & 96(4) & 102(4) & 83.3 \\ 
200 & 270.0 & 1.010(5) & 1.032 & 87(3) & 95(3) & 86.0 \\ 
200 & 290.0 & 0.902(7) & 0.951 & 91(3) & 99(3) & 90.4 \\
200 & 310.0 & 0.79(1) & 0.856 & 109(5) & 118(4) & 97.7 \\
800 & 250.0 & 1.216(2) & 1.211 & 76(3) & 82(2) & 74.6 \\ 
800 & 270.0 & 1.159(2) & 1.165 & 71(2) & 78(2) & 73.7 \\ 
800 & 290.0 & 1.107(2) & 1.118 & 72(2) & 80(2) & 73.0 \\
800 & 310.0 & 1.057(2) & 1.073 & 68(2) & 77(2) & 72.4 \\ 
\end{tabular}
\end{ruledtabular}
\footnotetext[1]{Ref. \onlinecite{webbookfluid}}
\end{table}

\begin{figure}
\includegraphics{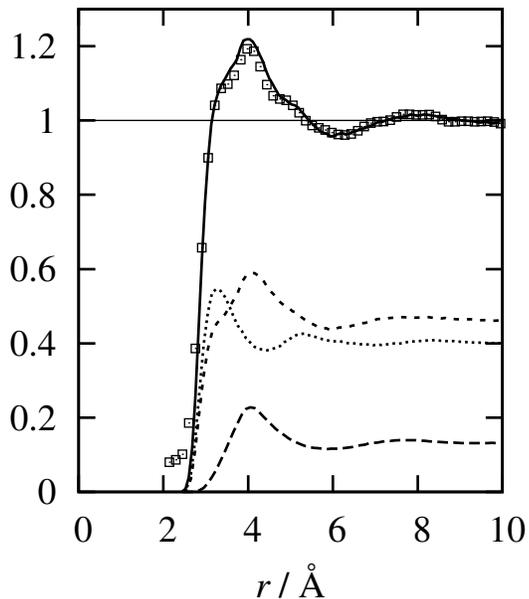}
\caption{The individual atomic PDFs, $g(r)$, of Eq.~(\ref{eq:scatter}) for
carbon-carbon (long-dashed line), carbon-oxygen (short-dashed line) and
oxygen-oxygen (dotted line) at 312 K and 0.83 g / cm$^3$ for the GCPCDO IP as
well as their superposition (full line), weighted by occurrence and scattering
propensity, $G(r)$.  Squares are experimental results from Ref.
\onlinecite{cipriani98}. For clarity, no intramolecular contributions are shown
for the computed results.}
\label{fig:rdf}
\end{figure}

\begin{figure}
\includegraphics{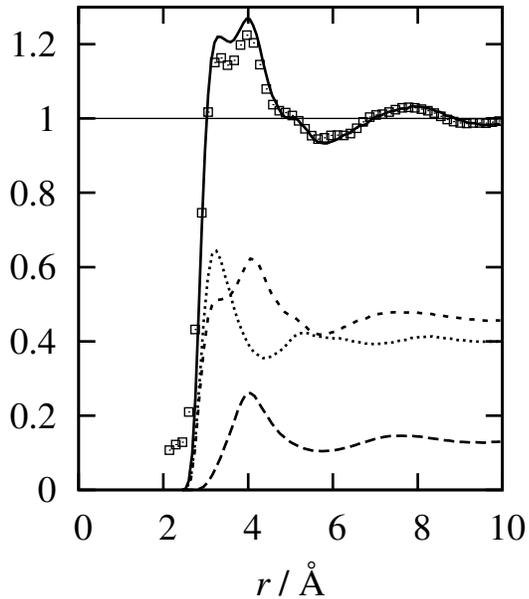}
\caption{Same as Figure \ref{fig:rdf} but for 1.09 g / cm$^3$ and 240 K.}
\label{fig:rdf2}
\end{figure}

No IP for the fluid can be deemed satisfactory if unable to reproduce the
experimental fluid structure. Accordingly, at the density and temperature of
the neutron-diffraction experiments of Cipriani {\em et al.},\cite{cipriani98}
the atomic pair distribution functions (PDF), $g(r)$, have been computed, and
these are shown in Figures \ref{fig:rdf} and \ref{fig:rdf2} for two different
thermodynamic states.  What is experimentally ascertained, however, is not the
individual, atomically resolved $g(r)$, but the superimposed effect from all
atomic scatterers. Taking account not only of the four times greater abundance
of C-O and O-O vectors than of C-C vectors between molecules, but also of the
different propensity toward neutron scattering, the following formula was used
to calculate the superimposed PDF, \cite{vantricht84}
\begin{eqnarray}
G(r) & = & 0.403 g_\mathrm{OO}(r) + 0.464 g_\mathrm{CO}(r) + 0.133
g_\mathrm{CC}(r) \label{eq:scatter}
\end{eqnarray}
In terms of the position and number of peaks, these $G(r)$ are seen to be in
reasonable-to-good agreement with the experimental results.


At 240 K and 1.09 g / cm$^3$, the carbon-oxygen distribution function clearly
shows more structure at short range, than at 312 K and 0.83 g / cm$^3$, where
the lack of orientational correlation in the fluid is also apparent in how
quickly the atomic carbon-oxygen and oxygen-oxygen PDFs decay to unity. Still,
however, the carbon-carbon PDF exhibits a slight peak at around 7.5~\AA\
indicating a weak second coordination shell, albeit of random order in
molecular orientation. The PDFs indicate that the fluid is slightly
overstructured at the higher density, where the first peak is overestimated.
The better agreement for the computed PDF at the low density is most likely due
to the overall lesser contributions from many-body effects at this density. For
the pair potential used in Ref.  \onlinecite{cipriani98}, the first peak is
overestimated at both thermodynamics states. Last, it should be pointed out
that in the fluid with flexible bonds, a general broadening of the peak
structure is expected.  This effect is at least responsible for the deviation
seen at the very short distances, where the internal scattering vectors
contribute. In the rigid model, they are $\delta$-functions and have been
omitted for clarity.

\subsection{Clusters}
Having established the weakness of the GCPCDO model primarily in its
$B_4(T)$, {\em i. e.} four-body interaction, we turn to properties of the IP
for which only three bodies contribute. Clusters like these offer excellent
tests of the model, due to the availability of quantum-mechanical reference
calculations. It also allows us to pinpoint more clearly the role of many-body
effects in the interaction potential.

\subsubsection{Dimer and trimers}
\begin{figure}
\includegraphics{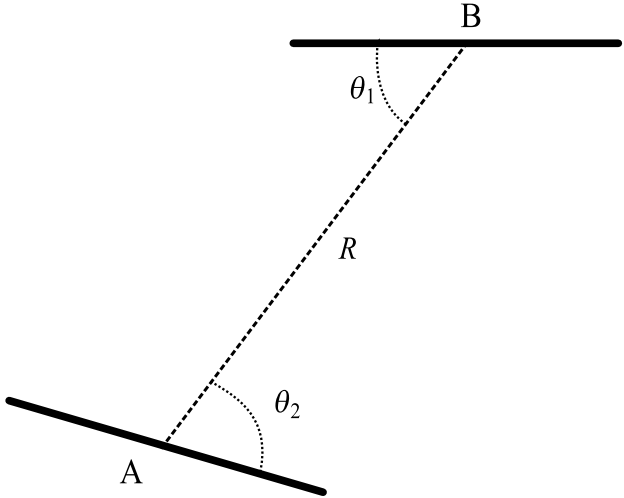}
\caption{Schematic illustration of the dimer with definitions of angles and
distances. Each molecule is represented by a stick. All atoms lie in the same
plane.}
\label{fig:dimer}
\end{figure}

Both experiment \cite{barton78, walsh87} and {\em ab initio} simulation
\cite{steinebrunner98, bukowski99, bock00} agree that the equilibrium dimer
structure is of $C_{2h}$ symmetry, with the {\em ab initio} simulations
indicating that there is a saddle point of $C_{2v}$ symmetry. The GCPCDO
IP reproduces these two dimer states very well, as shown in Table
\ref{tab:dimer}. As for the geometry of the dimer configurations (see Figure
\ref{fig:dimer}), it is neither better nor worse than the simpler ZD
IP,\cite{zhang05} but when it comes to the binding energy of the two
states, it is markedly superior when judged against the {\em ab initio}
IPs: the binding energy is underestimated by less than 4\% for the minimum
and overestimated by less than 1\% for the saddle-point, whereas the ZD IP
underestimates the binding energy in both cases by about 15--20\%. This gross
underestimation of the dimer binding energy helps explain the overestimation of
$B_2(T)$ for the ZD model ({\em vide supra}) but is nevertheless surprising
because for non-polarizable potentials the general trend is an overestimation
of dimer binding energies.

\begin{table}
\caption{Equilibrium geometry and well-depth energy of the CO$_2$ dimer for the
GCPCDO model, the ZD model, and BUK, the angular {\em ab initio} potential
energy surface of Bukowski {\em et al}.\cite{bukowski99} $U$ refers to the
potential well-depth at the specific conformation. See Figure \ref{fig:dimer}
for the definitions of the geometric quantities.}
\label{tab:dimer}
\begin{ruledtabular}
\begin{tabular}{r l r r r r} 
 & & GCPCDO & ZD & BUK\footnote{Ref. \onlinecite{bukowski99}} &
 Exp.\footnote{Ref. \onlinecite{walsh87}} \\ 
\multicolumn{2}{l}{Minimum ($C_{2h}$)} & & & & \\
 & $U$ / K                     & $-675.4$ & $-548.7$ & $-696.8$ & \\ 
 & $\theta_1, \theta_2$ / deg  & $55.5$   & $56.5$   & $59.0$   & $57.96$ \\ 
 & $R$ / \AA                   & $3.64$   & $3.64$   & $3.54$   & $3.60$ \\ 
\multicolumn{2}{l}{Saddle-point ($C_{2v}$)} & & & &  \\
 & $U$ / K & $-596.6$ & $-508.7$ & $-593.2$ & \\ 
 & $\theta_1$ / deg & $90.0$ & $90.0$ & $90.0$ & \\
 & $\theta_2$ / deg & $0.0$ & $0.0$ & $0.0$ & \\
 & $R$ / \AA & $4.16$ & $4.17$ & $4.14$ & \\ 
\end{tabular}
\end{ruledtabular}
\end{table}

\begin{figure}
\includegraphics{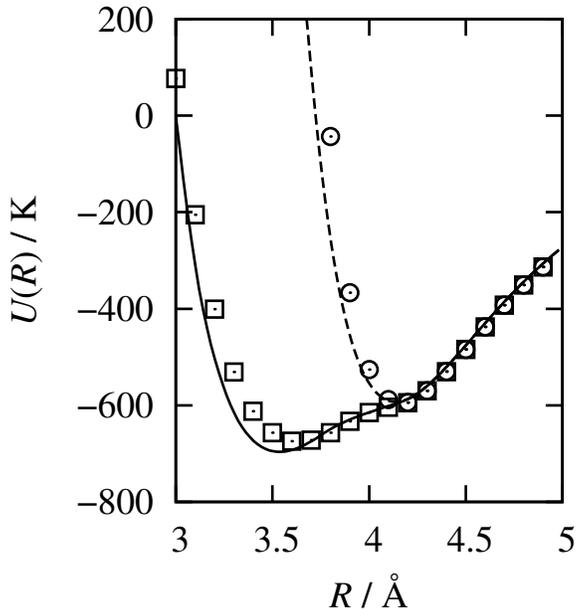}
\caption{Dependence of energy of interaction, $U(R)$, on separation, $R$, for
two different paths of approach. Squares are for the energy minimal path of the
GCPCDO IP and the full line is the analogous result for the BUK IP. Circles are
for the $C_{2v}$ conformation of the GCPCDO dimer, and the dashed line is for
the BUK result.} \label{fig:dissoc} 
\end{figure}

Because the GCPCDO model is so successful at capturing the dimer binding
energy, it is interesting to test it across a broader range of conformations.
Hence, I show in Figure \ref{fig:dissoc} the potential energy function of this
model compared to the accurate data of the angular fit of the symmetry-adapted
perturbation theory\cite{jeziorski94} calculations of Bukowski {\em et al.}
,\cite{bukowski99} the BUK IP, for radial displacements.  Both the energy
minimal pathway separating the two molecules and the path along the $C_{2v}$
transition state are shown. For the energy optimal pathway, the optimized
angles of the two molecules at each separation are shown in Table
\ref{tab:dissoc}.
The two molecules, for both the GCPCDO and BUK IPs, prefer a slipped-parallel
conformation at close range, but eventually prefer the T-shaped geometry of the
minimum at long range. The transition is noticeable as a slight trough in the
dissociation curve at around 4 \AA\ and is quicker for the BUK IP with an
earlier onset. 

Another interesting property of the model, which cannot be answered by the
BUK IP, is the total dipole of the $C_{2v}$ configuration. Because
the electric field gradients are very inhomogeneous close to the molecule, it
might be suspected that only allowing the center atom to polarize is
artificially deflating the induced dipole, and like this introducing errors in
the short-range interaction. For GCPCDO, the dipole is predicted to be 0.19 D
at the transition-state separation. Running calculations with Gaussian 03
,\cite{gaussian03} the suspicion is confirmed as calculations of the dipole
moment in the identical configuration at the CISD/aug-cc-pVDZ level of theory,
predict a dipole moment of 0.21 D. The dipole moment is hence underestimated by
10\% in the transition state configuration. 
\begin{table}
\caption{The optimum values of the angles $\theta_1$ and $\theta_2$ as a
function of the separation $R$ between two molecules as predicted by the
GCPCDO IP and the BUK IP. The angles are defined geometrically
in Figure \ref{fig:dimer}.}
\label{tab:dissoc}
\begin{ruledtabular}
\begin{tabular}{l r r r r} 
 & \multicolumn{2}{l}{GCPCDO} & \multicolumn{2}{l}{BUK} \\
 $R$ / \AA & $\theta_1$ / deg & $\theta_2$ / deg& $\theta_1$ / deg & $\theta_2$
 / deg \\ \hline
 3.0       & $71.7$ & $71.7$ & $66.4$ & $66.4$ \\ 
 3.3       & $62.3$ & $62.3$ & $62.2$ & $62.2$ \\ 
 3.5       & $58.0$ & $58.0$ & $59.5$ & $59.5$ \\ 
 3.7       & $54.5$ & $54.5$ & $57.0$ & $57.0$ \\ 
 3.8       & $53.0$ & $53.0$ & $46.1$ & $65.2$ \\ 
 3.9       & $42.1$ & $61.3$ & $37.1$ & $71.1$  \\
 4.0       & $30.8$ & $70.2$ & $28.6$ & $75.9$ \\
 4.1       & $19.6$ & $77.9$ & $18.9$ & $80.9$ \\
 4.2       & $3.7$  & $88.0$ & $0.0$  & $90.0$ \\
 4.5       & $0.0$  & $90.0$ & $0.0$  & $90.0$ \\ 
\end{tabular}
\end{ruledtabular}
\end{table}

\begin{figure}
\includegraphics{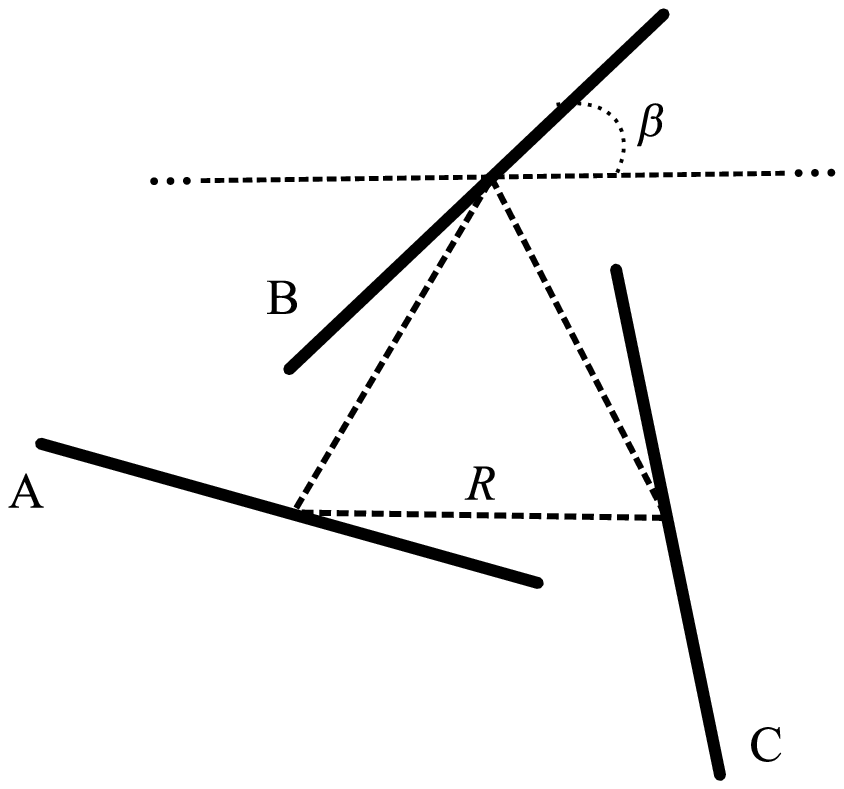}
\caption{Schematic illustration of the $C_{3h}$ trimer. All atoms lie in the
same plane. Molecules A, B and C are all identical by symmetry.}
\label{fig:tr2}
\end{figure}

\begin{figure}
\includegraphics{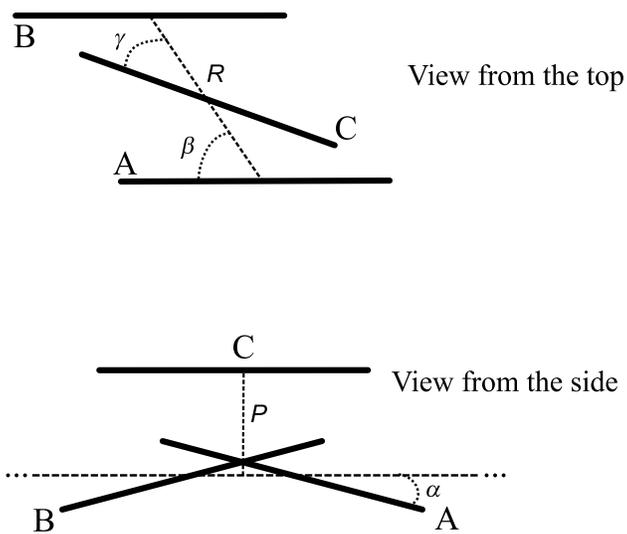}
\caption{Schematic illustration of the $C_2$ trimer. Molecules A and B are
identical by symmetry and the distance between their centers is $R$. The
perpendicular distance from the center of the line joining their centers to
the center of molecule C is $P$.}
\label{fig:tr1}
\end{figure}

Because of its many-body nature, it is interesting to test the GCPCDO IP
on the simplest cluster for which many-body effects contribute, {\em i. e.} the
trimer. Consequently, energy minimized structures of the trimer have been
located. The two most stable conformations are shown in Figures \ref{fig:tr2}
and \ref{fig:tr1}; the specific data on each are summarized in Table
\ref{tab:trimers}. Both of these trimer conformations have been observed
spectroscopically, with the planar $C_{3h}$ conformation being slightly more
abundant.\cite{weida95,weida96} In terms of relative energies, however, neither
of GCPCDO or BUK  predict the right ordering, but there are two general
remarks to be made. The first one is that the inclusion of many-body
effects, clearly levels the difference between the two states, the difference
in the GCPCDO prediction being less than $0.1$ K.\footnote{The precise value
of this difference is uncertain because of the numerical minimization involved,
but my program code indicates that the $C_2$ conformation lies $0.03$ K above
the $C_{3h}$ one.} That many-body effects would alleviate the problem was
hinted at already by Bukowski and coworkers \cite{bukowski99} in their
discussion of this problem and they argued using single-point calculation from
higher-order quantum chemical theory that this was the case.  Second, the
possible role of the zero-point vibrational quanta has to be kept in mind.  A
first-order estimate of this effect is provided by the harmonic zero-point
energy. Indeed, as indicated by Bukowski and coworkers,\cite{bukowski99}
inclusion of this energy for the BUK IP decreases the difference between the
states. Carrying out the same analysis for the GCPCDO IP, however, we find that
theory is brought into qualitative agreement with observation. As discussed by
Bukowski and collaborators,\cite{bukowski99} the harmonic approximation is
very strained in the CO$_2$ trimer but correct evaluation of the zero-point
vibrational energy necessitates a numerical solution of the Schr\"odinger
equation. Future code development will remedy this situation.  Even so, I
believe that these results are indicative of the qualities of the GCPCDO IP. 

\begin{table}
\caption{Energy well depth, $U$, equilibrium geometry and harmonic zero-point
vibrational energy, $E_0$, for the two most stable trimers predicted by the
GCPCDO IP and the BUK IP. Because of uncertainties in the numerical Hessian,
$E_0$ is rounded for the GCPCDO. The geometric variables are defined
in Figure \ref{fig:tr1} for the $C_2$ minimum and in Figure \ref{fig:tr2} for
the $C_{3h}$ minimum.} \label{tab:trimers}
\begin{ruledtabular}
\begin{tabular}{r l r r} 
 & & GCPCDO & BUK\footnote{Ref. \onlinecite{bukowski99}} \\ \hline
\multicolumn{2}{l}{$C_2$ minimum} & & \\
 & $U$ / K &      $-1849.4$ & $-1889.8$ \\
 & $E_0$ / K &      $359$   & $343.6$ \\
 & $R$ / \AA &      $3.73$  & $3.60$ \\
 & $P$ / \AA &      $2.96$  & $2.89$ \\
 & $\alpha$ / deg & $12.1$  & $10.9$ \\ 
 & $\beta$ / deg &  $53.7$  & $53.4$ \\ 
 & $\gamma$ / deg & $152.8$ & $157.6$ \\ 
\multicolumn{2}{l}{$C_{3h}$ minimum}  & & \\ 
 & $U$ / K   &   $-1849.4$ & $-1819.6$ \\ 
 & $E_0$ / K &     $318$ &  $304.2$ \\
 & $R$ / \AA &     $4.07$  &  $4.04$ \\
 & $\beta$ / deg & $40.9$  &  $39.3$ \\ 
\end{tabular}
\end{ruledtabular}
\end{table}

\subsubsection{Water complexes}
Because of its transparent physical form, the GCPCDO IP can, through the
adoption of suitable ``combining rules'', be interfaced with other IPs.
As a first test of the feasibility of this approach, I have calculated the
binding energy and molecular geometry of the \ce{[H2O-CO2]} complex using the
successful GCPM water \cite{paricaud05} for the water moiety. In addition to the
combining rules of Eqs~(\ref{eqn:mix1}) and (\ref{eqn:mix2}), the
$\gamma$-parameters were calculated as
\begin{equation}
\gamma_{xy}  =  \frac 1 2 (\gamma_x + \gamma_y)
\end{equation}
For comparison purposes, the complexation of ZD CO$_2$ and TIP3P
water\cite{jorgensen83} serves as an indicator of the effect of neglected
polarization. These results are summarized in Table \ref{tab:cmplx}. It is
important to point out that the potential energy surface of this complex is
very shallow near the minimum so precision is difficult, and two different {\em
ab initio} minimum energy geometries have been reported in the literature.
Danten and collaborators \cite{danten05} find that the energy minimum has $C_s$
symmetry
from counterpoise-corrected MP2 calculations with the aug-cc-pVTZ basis set.
However, this is contested by both experimental \cite{peterson84, tso85} and
more high-level {\em ab initio} counterpoise-corrected calculations at the
CCSD(T) level of theory and the same basis set which predict $C_{2v}$ symmetry
of the complex.\cite{garden06} It is interesting to note that the MP2 level of
theory often overestimates correlation effects, consistent with the fact that
simple pair potentials, such as the ZD \cite{zhang05} and TIP3P
\cite{jorgensen83} IPs under the Lorentz-Berthelot mixing rules, predict
a $C_{2v}$ minimum for the complex. The intermixing of the GCPCDO IP
with GCPM water \cite{paricaud05} produces two, equivalent, global minima of
$C_s$ symmetry, but the difference in energy between them and the transition
state of $C_{2v}$ symmetry connecting them is only about 7 K, {\em i. e.} less
than 0.1\% of the total interaction energy and should not be taken as great
drawback of the model. 

It is a greater error that the total binding energy is
underestimated by about 25\%, a fault shared by the ZD/TIP3P combination as
well. This is surprising, because in general non-polarizable IPs parametrized
against bulk properties tend to overestimate cluster binding energies. The
explanation can partly be found in the rigid geometries of the GCPCDO/GCPM and
ZD/TIP3P moieties. The results by Danten and coworkers \cite{danten05} indicate
that the CO$_2$ molecule is bent slightly upon complexation, thus further
inducing a dipole moment which increases the energy of interaction. However,
the greater part of the explanation is to be found in the approximation that
only the central atom polarizes. A QM calculation at the CISD/aug-cc-pVDZ level
of theory for the $C_s$ minimum of the GCPM/GCPCDO complex indicate that while
the total dipole moment of 2.19~D is in good agreement with the accurate value
of 2.21 D, the individual components are poorly reproduced.
\footnote{It must be remembered, however, that a large part of this dipole
moment is due to the ground-state charge distribution of the water molecule,
captured already by the partial charges in the GCPM water.} 
The QM calculation indicates that for the complex as a whole, the electric
dipole moment along the $C_2$ axis of the water molecule is 2.16 D, and that
the perpendicular component is 0.465 D in this particular configuration. The
GCPM/GCPCDO pairing, however, indicates 2.19 D and 0.14 D, respectively, for
these components. Further discrepancy is expected at the closer range indicated
as the equilibrium distance by the {\em ab initio} calculations. In summary, as
far as geometry and energy are concerned, the predictions of the GCPCDO/GCPM
complex can hardly be considered an improvement over the simple empirical IPs,
compared to {\em ab initio} calculations.
\begin{table}
\caption{Minimum energy, $U$, and geometry of the \ce{[CO2-H2O]} complex as
predicted by the mixing of GCPCDO and GCPM polarizable IPs, the ZD and
TIP3P non-polarizable IPs, or {\em ab initio} calculation by Danten {\em
et al.} at the MP2/aug-cc-pVTZ level of theory, or by Garden {\em et al.} at
the CCSD(T)/aug-cc-pVTZ level of theory. The binding energy of the $C_{2v}$
transition state predicted by the GCPCDO model is reported for completeness. For
the definition of the geometry, see Figure \ref{fig:cmplx}.}
\label{tab:cmplx}
\begin{ruledtabular}
\begin{tabular}{l r r r r r} 
 & \multicolumn{2}{l}{GCPCDO/GCPM} & ZD/TIP3P &
 Danten\footnote{Ref. \onlinecite{danten05}} & Garden\footnote{Ref.
 \onlinecite{garden06}} \\ 
 & $C_s$ minimum  & $C_{2v}$ tr. state &          &        &        \\ \hline
$U$ / K & $-1036.4$  & $-1029.5$ & $-1048.1$  & $-1308.4$ & $-1358.7$  \\
$R$ / \AA & $3.06$  & $3.06$  & $2.93$  & $2.77$ & $2.81$ \\
$\phi$ / deg & $17.0$  & $0.0$ &  $0.0$   &  $13.9$ & $0.0$ \\ 
$\alpha$ / deg & $87.4$ & $0.0$ &  $0.0$   &  $88.0$ & $0.0$ \\ 
\end{tabular}
\end{ruledtabular}
\end{table}
\begin{figure}
\includegraphics{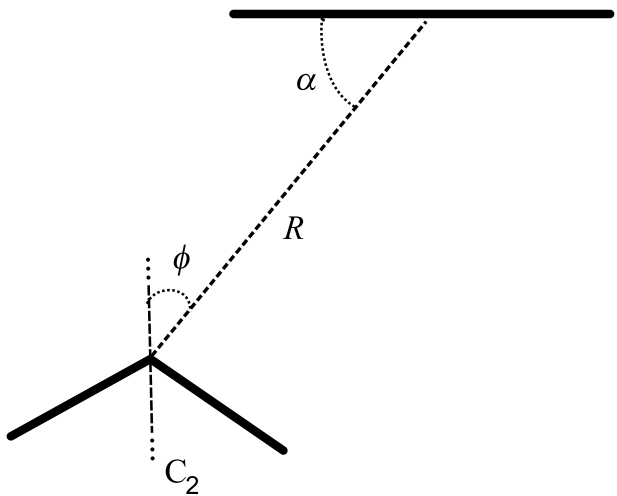}
\caption{Schematic illustration of the structure of the monohydrate complex.
The $C_2$ axis of the water molecule is marked. All atoms lie in the same
plane.}
\label{fig:cmplx}
\end{figure}

\begin{table}
\caption{Characteristics of the global minima of the \ce{[(CO2)2-H2O]}
complex predicted by the GCPCDO/GCPM IPs, the ZD/TIP3P IPs or
the {\em ab initio} results of Danten {\em et al.} at the MP2/aug-cc-pVTZ level
of theory. The oxygen of the water molecule is placed at the origin of a
Cartesian coordinate system in which the $y$-axis coincides with the $C_2$ axis
of the water molecule with the positive direction pointing toward the hydrogen
atoms, the $z$-axis is normal to the molecular plane and the $x$-axis is
orthogonal to the $y$-- and $z$-axes. The centers of mass of the CO$_2$
molecules
are given with respect to this coordinate system, and the orientation of each
molecule is expressed in the Euler angles $\alpha$, $\beta$ and $\gamma$ which
denote counterclockwise rotation around the $x$--, $y$-- and $z$-axes,
respectively. $U$ is the energy of that conformation. Indices $1$ and $2$
denote the two CO$_2$ molecules. The work in Ref. \onlinecite{danten05} used
flexible molecules. The coordinates reported here are rounded over these
differences.}
\label{tab:cmplx2}
\begin{ruledtabular}
\begin{tabular}{l r r r} 
  & GCPCDO/GCPM & ZD/TIP3P & Danten\footnote{Ref. \onlinecite{danten05}}  \\
  \hline
  Point group & $C_s$ & $C_2$ & $C_s$ \\
  $U$ / K & $-2738.5$  & $-5074.0$ & $-2918.7$  \\
  $x_1$ / \AA & $0.00$ & $-1.43$   & $0.00$  \\
  $y_1$ / \AA & $-2.61$ & $-2.11$  & $-2.52$   \\
  $z_1$ / \AA & $1.58$   & $-0.85$ & $1.11$    \\
  $x_2$ / \AA & $0.00$ & $1.43$ & $0.00$    \\
  $y_2$ / \AA & $0.79$  & $-2.11$  & $0.42$   \\
  $z_2$ / \AA & $3.80$   & $0.85$ & $3.80$    \\
  $\alpha_1$ / deg & $53.7$ & $-62.8$  &  $61.8$         \\ 
  $\beta_1$ / deg & $90.0$   & $86.0$  & $90.0$          \\ 
  $\gamma_1$ / deg & $90.0$   & $-82.3$  & $90.0$       \\ 
  $\alpha_2$ / deg & $-69.1$    & $62.8$ & $-55.0$      \\ 
  $\beta_2$ / deg & $90.0$ &   $86.0$ &  $90.0$        \\ 
  $\gamma_2$ / deg & $90.0$   & $82.3$ & $90.0$        \\ 
\end{tabular}
\end{ruledtabular}
\end{table}

Despite this small short-coming of the predictions for the \ce{[CO2-H2O]}
complex, it is still worthwhile to consider the \ce{[(CO2)2-H2O]}
complex, also studied by Danten and coworkers,\cite{danten05} because here the
true many-body interactions start to play a role. Moreover, contrary to the
case of the CO$_2$ trimer, where none of the moieties is dipolar, the water
molecule carries a substantial dipole moment and the electronic induction
effects are expected to be more pronounced. It must be pointed out that despite
this being a three-body system, no Axilrod-Teller potential has been applied.
The reason is that while GCPM water has a known polarizability, it has no
Axilrod-Teller coefficient. Rather than impose one on the model, I have decided
to judge it fairly according to its own merits, and these exclude a three-body
dispersion interaction.  The results are shown in Table \ref{tab:cmplx2}. The
agreement is very good in terms of energy. Contrary to the \ce{[CO2-H2O]}
complex, for the \ce{[(CO2)2-H2O]} complex,  there is no appreciable
underestimation of the binding energy. This can probably be attributed to the
larger fraction of CO$_2$-CO$_2$ interactions in this complex.  Moreover, both
Danten {\em et al.} \cite{danten05} and I find that the minimum is of $C_s$
symmetry, but for the ZD\cite{zhang05}/TIP3P\cite{jorgensen83} model, the
global minimum is of $C_2$ symmetry in a conformation reminiscent of the $C_2$
trimer. A local minimum of $C_2$ symmetry is predicted by the
GCPCDO/GCPM\cite{paricaud05} model at about $208$ K above the global minimum.
Clearly, polarization changes the relative stability of these two conformations
in favor of the $C_s$ one and this is captured both by the MP2 calculations of
Danten {\em et al.} \cite{danten05} and by the present work. The peculiarities
of the global minima predicted by the IPs are given in Table \ref{tab:cmplx2}.
It is very clear that the many-body effects are responsible for the altered
symmetry of the equilibrium structure. Also, the binding energy is vastly
overestimated by the non-polarizable IP pair. 

\section{Concluding remarks}
\label{sec3}
A new, polarizable IP for CO$_2$ has been introduced and shown to be in
excellent agreement dimer properties and excellent-to-passable agreement for the
bulk phase. Classical non-polarizable IPs that reproduce bulk phase properties
well have been shown inadequate for the description of $B_2(T)$ and also
$B_3(T)$. That many-body effects should not be ignored for the CO$_2$ molecule
is evidenced by the qualitative experimental agreement that is achieved for the
stability of the two trimer conformations, something which not even the {\em ab
initio} potential energy surface of Bukowski {\em et al.} \cite{bukowski99}
manages.  Further corroboration of the model is provided by calculations on the
water complexes, where especially the \ce{[(CO2)2-H2O]} complex is in good
agreement with {\em ab initio} calculation at the MP2/cc-aug-pVTZ level of
theory.\cite{danten05} Absence of polarization changes the symmetry of this
complex.   

A tough test for all of the molecular CO$_2$ potentials investigated is the
prediction of the virial coefficients. The GCPCDO model handles $B_2(T)$ and
$B_3(T)$, but only because of design, and fails at $B_4(T)$ and up. This can
still be considered an improvement over the classical, non-polarizable models
for which most probably none of the virial coefficients beyond the ideal gas
term are in agreement with experiment. Clearly, the interaction among CO$_2$
molecules is more complicated than a simple pairwise sum over atomic charges
and Lennard-Jones terms. However, it is also more complicated than
self-consistent solution of induced dipoles and triple-dipole dispersion
interaction.  Nevertheless, for systems of three bodies or less, it seems to be
highly satisfactory, meaning that the remaining errors relate to many-body
effects beyond the third. On the precise causes of the remaining errors in the
IP can only be speculated and future computer experiments may provide the
answer to what the mechanisms are.

It is in this light that it must be kept in mind that even if in comparison
with experiment, the GCPCDO model, with a few exceptions, rests more on
qualitative concordance than on many digits of precision, this is the
first many-body molecular IP for the CO$_2$ molecule to be developed and that
many areas of inquiry remain to explore. One obvious and immediate improvement
to the model would be to distribute the induced dipole over all atoms for a
better short-range description of the induced electrostatics.

The Fortran 90 source code for the GCPCDO energy subroutines is available upon
request.

\begin{acknowledgments}
The simulations were performed on the C3SE computing resources. The author also
wishes to express his gratitude to Prof. Robert Bukowski for granting him the
BUK subroutines and to Prof. Sture Nordholm for insightful comments on the
manuscript. Serious errors were noted by the anonymous reviewers and the author
is highly indebted to both.
\end{acknowledgments}

%\bibliography{patron}


\end{document}